\documentclass[amsmath,aps,prstab,longbibliography,preprint]{revtex4-1}

\usepackage{graphicx} 
\usepackage{hyperref}
\usepackage{amsmath}
\usepackage{slashed}

\begin{document}
\title{Electron dynamics in grating-type dielectric laser accelerators:
particle transfer function, generalized acceleration/deflection gradients and Panofsky-Wenzel theorem}

\author{Andrzej Szczepkowicz}

\address{Institute of Experimental Physics, University of Wroclaw, Plac Maksa Borna 9, 50-204 Wroclaw, Poland}

\date{\today}

\begin{abstract}
The notions of acceleration gradient and deflection gradient are generalized to phasor quantities (complex-valued functions) in the context of dielectric laser acceleration (DLA). It is shown that the electromagnetic forces imparted on a near-resonant particle traversing a unit cell of a grating-type DLA can be conveniently described by generalized acceleration and deflection gradients. A~simple formulation of the Panofsky-Wenzel theorem in terms of the generalized gradients is given. It is shown that all particle transfer properties of a DLA unit cell can be derived from a single, complex-valued function, the generalized acceleration gradient.
\end{abstract}


\maketitle

\section{Introduction}\label{sect-intro}

Dielectric laser acceleration (DLA) is one of the possible techniques for future compact accelerators \cite{Peraltaetal2013,BreuerHommelhoff2013,dla2014,dla2016,WoottonEngland2017} (see also Rasmus Ischebeck's invited lecture at this conference). Ongoing conceptual, design, fabrication and experimental work on DLA is accompanied by simulations. Electron trajectories in DLA can be calculated using a variety of approaches, ranging from computationally demanding particle-in-cell (PIC) simulations to simple tracking schemes implemented in general-purpose calculation tools. Recently, three simple particle tracking approaches were described in 
\cite{NiedermayerBoine-Frankenheim2017,NiedermayerEgenolf2017},
\cite{KuropkaAssman2017,MayetAssman2017}, 
and \cite{Szczepkowicz2017}. The three schemes analyze the transfer properties of dielectric unit cell and take advantage of the quasi-periodicity of DLA structures.

The present short communication formulates a natural generalization of the concept of acceleration and deflection gradients \cite{Plettner2007, PlettnerByer2008,LeedlePease2015, WeiJamison2017} for near-resonant particles (Sect.~\ref{sect-gg}), gives the relationship between the generalized gradients and the transfer function \cite{Szczepkowicz2017} (Sect.~\ref{sect-tfsa}), and formulates the Panofsky-Wenzel theorem \cite{PanofskyWenzel1956, TremaineRosenzweig1997, Chao2002} in terms of gradients (Sect.~\ref{sect-pw}; the theorem was already used in a similar context in Ref.~\cite{NiedermayerEgenolf2017}; see also slide 8 of Ref.~\cite{Hommelhoff2014}).
The results are summarized in Sect.~\ref{sect-s}.

\section{Generalized acceleration and deflection gradients}\label{sect-gg}

\begin{figure*}
	\centering
	\includegraphics[scale=1]{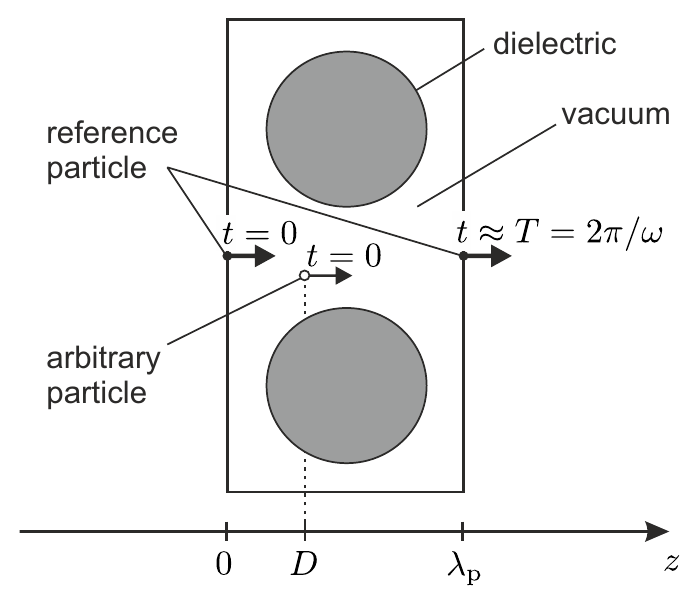}%
	\caption%
	{%
		\label{fig-unit-cell}%
	    Example DLA unit cell \cite{LeedleCeballos2015,Szczepkowicz2017}.
	    For a near-resonant particle, time of flight is approximately equal to the optical period. For the reference particle $z=c\beta t$ and $\Phi=0$. For an arbitrary particle $z=c\beta t+D$ and $\Phi=-k_\text p D$, where $\Phi$ is the optical phase when the particle enters the unit cell.
	    At the entrance of the cell $z=0$, $t=(0-D)/(c\beta)$ and
	    $\exp(i\omega t)=\exp[i k_\text p (-D)]=\exp(i\Phi).$
	}
\end{figure*}

The accelaration gradient $G_\textbf{acc}$ (= $G_z$) and
deflection gradient $G_\textbf{defl}$ (= $G_x, G_y$, alternative names: accelerating/deflecting gradients) are defined as the accelerating and deflecting fields (= force per unit charge) acting on a particle,
averaged over one DLA unit cell 
\cite{Plettner2007, PlettnerByer2008,LeedlePease2015, WeiJamison2017}, see Fig.~\ref{fig-unit-cell}.
Using this definition and the expression for the
Lorentz force, the following expressions for gradients are obtained in SI units:
\begin{subequations}
	\label{eq-G}
\begin{align}
	 G_x&=\frac{1}{\lambda_\text p} \int_{0}^{\lambda_\text p}
[E_x(x,y,z,t(z))-c\beta B_y(x,y,z,t(z))]\, dz
	\\
	 G_y&=\frac{1}{\lambda_\text p}\int_{0}^{\lambda_\text p}
[E_y(x,y,z,t(z))+c\beta B_x(x,y,z,t(z))] \, dz
	\\
	 G_z&=\frac{1}{\lambda_\text p}\int_{0}^{\lambda_\text p}
	 E_z(x,y,z,t(z))\,  dz
\end{align}
\end{subequations}
where $\lambda_\text p$ denotes the length of the unit cell.
Note that $x$ and $y$ are constant under the integral
-- this is a \emph{zero angle approximation},
where the particle motion is in the
longitudinal direction $\hat z$ with no transverse
velocity components.
The gradient definitons (\ref{eq-G}) can be expressed in a compact, vector from:
\begin{equation}
	\label{eq-Gvec}
	\vec G=\frac{1}{\lambda_\text p} \int_{0}^{\lambda_\text p}
	(\vec E+c\beta \hat z \times \vec B)\, dz
\end{equation}
Note that  $c\beta \hat z \times \vec B$ is perpendicular to $\hat z$,
so the magnetic force does not appear in the expression for $G_z$ (Eq.~(\ref{eq-G}c)).

Let us now assume that the electromagnetic field is stationary (time harmonic):
$E_x(x,y,z,t)=\Re[\widetilde E_x(x,y,z) e^{i \omega_0 t}]$ etc., and that the particle is close to resonance with the electromagnetic field, so that the length of the DLA unit cell is traversed in time almost equal to the optical period $T$.
The optical phase under the integrals (\ref{eq-G})
changes from $\Phi$ (at the entrance of the cell) to approximately $\Phi+2\pi$ (at the exit),
so the gradients may be expressed as
\begin{subequations}
	\label{eq-G1}
	\begin{align}
	G_x(x,y;\Phi)&=\frac{1}{\lambda_\text p} \int_{0}^{\lambda_\text p}
	\Re\left[\left(\widetilde E_x(x,y,z)-c\beta\widetilde B_y(x,y,z) \right)e^{i\Phi}e^{i k_\text p z}\right] dz
	\label{eq-G1x}
	\\
	G_y(x,y;\Phi)&=\frac{1}{\lambda_\text p}\int_{0}^{\lambda_\text p}
	\Re\left[\left(\widetilde E_y(x,y,z)+c\beta\widetilde B_x(x,y,z) \right)e^{i\Phi}e^{i k_\text p z}\right] dz
	\label{eq-G2y}
	\\
	G_z(x,y;\Phi)&=\frac{1}{\lambda_\text p}\int_{0}^{\lambda_\text p}
	\Re\left[\widetilde E_z(x,y,z)\, e^{i\Phi}e^{i k_\text p z}\right] dz
	\label{eq-G3z}
	\end{align}
\end{subequations}
where $k_\text p=2\pi/\lambda_\text p$. Note that the gradients depend on the phase of the field at the moment when the particle enters the unit cell. This dependence is sinusoidal,
so it is natural to introduce here complex-valued gradients:
\begin{subequations}
	\label{eq-GG}
	\begin{align}
	\widetilde G_x(x,y)&=\frac{1}{\lambda_\text p} \int_{0}^{\lambda_\text p}
	\left(\widetilde E_x(x,y,z)-c\beta\widetilde B_y(x,y,z) \right)e^{i k_\text p z}\, dz
	\label{eq-GGx}
	\\
	\widetilde G_y(x,y)&=\frac{1}{\lambda_\text p}\int_{0}^{\lambda_\text p}
	\left(\widetilde E_y(x,y,z)+c\beta\widetilde B_x(x,y,z) \right)e^{i k_\text p z}\, dz
	\label{eq-GGy}
	\\
	\widetilde G_z(x,y)&=\frac{1}{\lambda_\text p}\int_{0}^{\lambda_\text p}
	\widetilde E_z(x,y,z)\, e^{i k_\text p z}\, dz
	\label{eq-GGz}
	\end{align}
\end{subequations}
The generalized gradients (\ref{eq-GG}) are complex functions which do not depend on $\Phi$; instead, the phase dependence
is encoded in their complex argument, and ordinary phase-dependent gradients can be calculated from 
\begin{equation}
G_j(x,y;\Phi)=\Re[e^{i\Phi}\widetilde G_j(x,y)]
\end{equation}
where $j=x,y,z$.
From Eqns.~(\ref{eq-GG}) it can be seen that the generalized gradient is the amplitude (Fourier coefficient) of the first spatial harmonic of the stationary force field in a DLA cell (see also Refs. \cite{Plettner2007} and \cite{NiedermayerEgenolf2017}). 

The momentum kicks received by the particle in a unit cell are, by the definition
of gradients, 
$\Delta p_j
=\int F_j dt
=q\langle F_j/q \rangle T
=q\, G_j(x,y;\Phi)T
=q\,\Re[e^{i\Phi}\widetilde G_j(x,y)]T$,
so
\begin{equation}
	\label{eq-tf-aux}
	\Delta p_j(x,y;\Phi)=(-e)\Re[e^{i\Phi}\widetilde G_j(x,y)]\frac{\lambda_\text p}{\beta c}
\end{equation}

\section{From gradients to the transfer function in the small-angle, near-resonant approximation}\label{sect-tfsa}

In Reference~\cite{Szczepkowicz2017}, the notion of a nonlinear, phase-dependent
transfer function $\cal R$ was used, inspired by the transfer matrix formalism
($\cal R$-matrix, see eg.\ \cite{HemsingStupakov2014}). From this moment, to avoid confusion, the nonlinear, phase-dependent transfer function will be denoted by $\slashed{\cal R}$.
The transfer function $\slashed{\cal R}$ maps the particle parameters at the entrance of a DLA cell to the parameters at the exit:
\begin{equation}
(x_1,x'_1,y_1,y'_1,\Phi_1,\delta_1)
\xrightarrow{\displaystyle\slashed{\cal R}}
(x_2,x'_2,y_2,y'_2,\Phi_2,\delta_2).
\end{equation} 
The first four parameters have a similar meaning as in the conventional $\cal R$-matrix theory, except that the ranges of $x$ and $y$ cover the whole aperture of the device
(in a RF accelerating cavity $x,y$ are deflections from the design trajectory which are much smaller than the aperture). The fifth parameter is $\Phi$, the phase of the electromagnetic field at the entrance/exit of the DLA cell. It is not a small parameter. 
The entry phase $\Phi$ is considered here a property of the particle, one of the parameters that characterizes the motion of the particle. From cell to cell, $e^{i\Phi}$ is a slowly changing particle parameter. The sixth parameter, $\delta$, denotes the particle's relative deviation from the reference momentum $p_0$ (in Ref.~\cite{Szczepkowicz2017} $p_0$ was fixed for the whole DLA element (consisting of 16 DLA cells), while in Ref.~\cite{NiedermayerEgenolf2017} the reference energy $W_0$ varies from cell to cell (for hundreds of cells) according to the design \emph{acceleration ramp}).

In the Appendix A of Ref \cite{Szczepkowicz2017}, the equations for the transfer function
were given in full form, without the assumption that $x'$ and $y'$ are small.
This full form could be relevant for the early stages of acceleration, but at
higher particle energies it is safe to assume $x',y'\ll 1$. Within this
\emph{small angle approximation}, the trajectory
deflection cosine $C=\frac{1}{\sqrt{{x'}^2+{y'}^2+1}}$ can be replaced by 1 in the transfer equations, and longitudinal velocity $\beta_z$ can be replaced by velocity $\beta$. Let us also assume that the particle is close to resonance, so the ,,momentum kicks'' 
$\Delta p_x, \Delta p_y, \Delta p_z$ can be calculated
from $\widetilde G_x, \widetilde G_y, \widetilde G_z$ as described in Sect.~\ref{sect-gg}.

With the above simplifications, the expressions for the transfer function are
\begin{subequations}\label{eq-tf}
	\begin{align}
	x_2&=x_1+x'_1\,\lambda_\text p\label{eq-tf-x}\\
	x'_2&=
	x'_1+\frac{\Delta p_x}{ p_0(1+\delta_1)}\label{eq-tf-xp}\\
	y_2&=y_1+y'_1\,\lambda_\text p\label{eq-tf-y}\\
	y'_2&=y'_1+\frac{\Delta p_y}{ p_0(1+\delta_1)}\label{eq-tf-yp}\\
	\Phi_2&=\Phi_1+\frac{\omega}{c\beta}\lambda_\text p\label{eq-tf-Phi}\\
	\delta_2&=\delta_1+\frac{\Delta p_z}{ p_0}\label{eq-tf-delta}
	\end{align}
\end{subequations}
where the momentum kicks
are calculated from gradients (Eq.~(\ref{eq-tf-aux}));
in the calculation of $\Phi$ (Eq.~(\ref{eq-tf-Phi})), the increase in electron's $\beta$ is taken into account (see also Appendix C of Ref.~\cite{Szczepkowicz2017}):
\begin{equation}
\beta=\frac{p_0(1+\delta_1)}{\sqrt{p_0^2(1+\delta_1)^2+m^2c^2}}\label{eq-tf-aux-betaz}
\end{equation}
In this approach the three-dimensional form of the unit cell is arbitrary, with no assumptions about symmetry. The stationary fields $\widetilde E_x,\widetilde E_y,\ldots$ and the gradients must be computed numerically.

Note that the scheme (\ref{eq-tf}) is a simple Euler numerical method; better accuracy can be obtained
using the symplectic-Euler method \cite{NiedermayerEgenolf2017}.

\section{The Panofsky-Wenzel theorem for DLA in terms of gradients}\label{sect-pw}

The longitudinal and transverse momentum transfer from the electromagnetic field to the particle are related
by the Panofsky-Wenzel theorem \cite{PanofskyWenzel1956,TremaineRosenzweig1997,Chao2002}.
The theorem was originally derived in the context of wakefields, and is almost always discussed in
this context. Although wakefields are not at present
significant for DLA experiments, the Panofsky-Wenzel theorem is important for DLA
(see slide 8 of Ref.~\cite{Hommelhoff2014} and Ref.~\cite{NiedermayerEgenolf2017}).
Previous publications which use the notions of acceleration and deflection gradients in the context of DLA \cite{Plettner2007, PlettnerByer2008, LeedlePease2015, WeiJamison2017}
do not mention the Panofsky-Wenzel theorem.
A recent paper by Niedermayer, Egenolf and Boine-Frankenheim \cite{NiedermayerEgenolf2017} is the first article to use the theorem in the context of DLA.
The theoretical framework in \cite{NiedermayerEgenolf2017}
does not  use the notions of acceleration/deflection gradient -- instead, $\Delta W(x,y;s)$ and $\Delta \vec p_\perp(x,y;s)$ are used, see Eqs~(1) and (12) in Ref.~\cite{NiedermayerEgenolf2017}. The purpose
of this Section is to formulate the implications
of the Panofsky-Wenzel theorem in terms of acceleration/deflection gradients 
$\widetilde G_x, \widetilde G_y, \widetilde G_z$, and thus link the theorem
with previous literature that uses acceleration and deflection gradients in the context of DLA \cite{Plettner2007, PlettnerByer2008, LeedlePease2015, WeiJamison2017}.

\begin{figure*}
	\centering
	\includegraphics[scale=1]{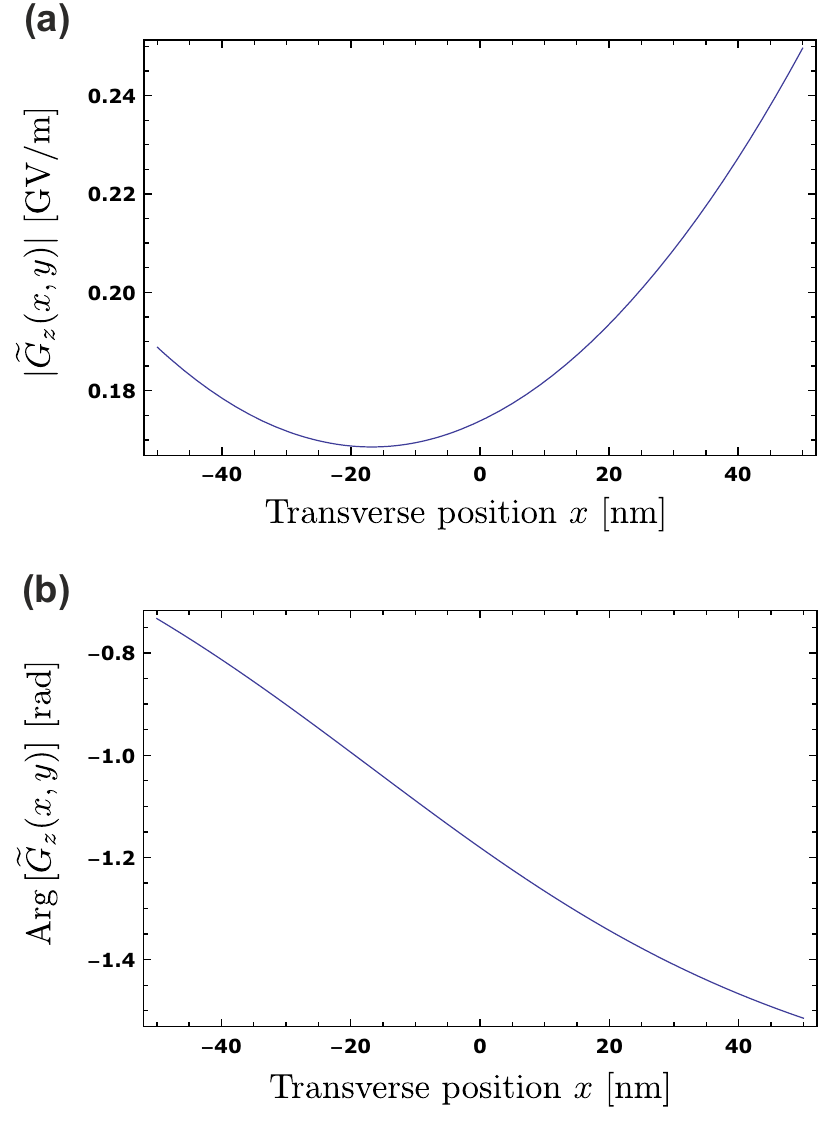}%
	\caption%
	{%
		\label{fig-acc}%
		All information about the dynamics of a near-resonant particle in a DLA unit cell
		can be derived from only two plots: (a)--modulus and (b)--argument, which show the acceleration gradient $\widetilde G_z(x,y)$. In this example $\widetilde G_z$ is calculated for a unit cell of Ref.~\cite{Szczepkowicz2017}.
	}
\end{figure*}

The Panofsky-Wenzel theorem states that
\begin{equation}
\label{eq-pw-vec}
\vec{\nabla}\times\Delta\vec p=\vec 0,
\end{equation}
where the momentum kick $\Delta\vec p$ is a function of $(x,y,D)$ and $\vec{\nabla}$ refers to taking derivative
with respect to coordinates $(x,y,D)$ \cite{Chao2002}, and the $D$ coordinate \cite{Chao2002} is defined
in Fig.~\ref{fig-unit-cell}. (Note that Ref.~\cite{NiedermayerEgenolf2017} uses the variable $s$ and $D=-s$.) 
The theorem is usually applied for the case where the electromagnetic field vanishes outside
the region of interest \cite{TremaineRosenzweig1997, Chao2002}, but it is also valid for
periodic boundary conditions \cite{NiedermayerEgenolf2017} of grating-type DLA.
Eq.~(\ref{eq-pw-vec})
is equivalent to
\begin{subequations}
	\label{eq-pw-xyz}
\begin{align}
	\partial_y \Delta p_z - \partial_D \Delta p_y & =0 \label{eq-pw-x}\\
	\partial_D \Delta p_x - \partial_x \Delta p_z & =0 \label{eq-pw-y}\\
	\partial_x \Delta p_y - \partial_y \Delta p_x & =0 \label{eq-pw-z}
\end{align}
\end{subequations}
Let us now pass from differentiation with respect do $D$ to differentiation with
respect to $\Phi$ (compare Sect.~\ref{sect-gg} and \ref{sect-tfsa});
$\Phi=-k_\text p D$, so
\begin{equation}
\frac{\partial}{\partial D} = 
\frac{d \Phi}{d D} \frac{\partial}{\partial \Phi} =
-k_\text p \frac{\partial}{\partial \Phi}
\end{equation}
Equation (\ref{eq-pw-y}) is equivalent to
\begin{subequations}
	\begin{align}
	-k_\text p \partial_\Phi \Delta p_x &= \partial_x \Delta p_z\\
	-k_\text p \partial_\Phi \Re[e^{i\Phi}\widetilde G_x]  
	&= \partial_x \Re[e^{i\Phi}\widetilde G_z]\\
	\Re[-k_\text p \partial_\Phi (e^{i\Phi}\widetilde G_x)]  
&= \Re[\partial_x( e^{i\Phi}\widetilde G_z)] \label{eq-RkdGx}
	\end{align}
\end{subequations}
Equation (\ref{eq-RkdGx}) hold for all $\Phi$, so
\begin{subequations}
\begin{align}
-k_\text p \partial_\Phi (e^{i\Phi}\widetilde G_x)
&= \partial_x (e^{i\Phi}\widetilde G_z)\\
-ik_\text p \widetilde G_x
&= \partial_x \widetilde G_z\\
\end{align}
\end{subequations}
So the $x$-deflection gradient can be calculated from the acceleration gradient
\begin{equation}
\widetilde G_x
= -\frac{1}{ik_\text p }\partial_x \widetilde G_z \label{ex-pw-x-fin}\\
\end{equation}
Similarily, Eq.~(\ref{eq-pw-x}) is equivalent to
\begin{equation}
\widetilde G_y
= -\frac{1}{ik_\text p }\partial_y \widetilde G_z \label{ex-pw-y-fin}\\
\end{equation}
The last two equations can be put into a more compact form
\begin{equation}
\label{eq-pw-g}
\vec{\widetilde G}_\perp
= -\frac{1}{ik_\text p }\vec{\nabla}_{\!\perp}  \widetilde G_z\\
\end{equation}
Equation (\ref{eq-pw-g}) is the Panofsky-Wenzel theorem for near-resonant particles in DLA, expressed in terms of acceleration and deflection gradients; no assumptions about
symmetry were made except for $z\rightarrow z+\lambda_\text p$ translational symmetry (\emph{grating-type} DLA). The theorem states that the deflection gradient can be calculated from the acceleration gradient, and
implies that the particle dynamics (both longitudinal and transverse) in a DLA unit cell can be described by a single function $\widetilde G_z(x,y)$, the generalized acceleration gradient, as shown in Fig.~\ref{fig-acc}. 

In the above considerations, Eq.~(\ref{eq-pw-z}) was not used. 
The equation does not impose additional constraints on the gradients,
because in light of (\ref{ex-pw-x-fin}) and (\ref{ex-pw-y-fin}) it is equivalent to
\begin{subequations}
	\begin{align}
		\partial_x \widetilde G_y &= \partial_y \widetilde G_x\\
		\partial_x \partial_y \widetilde G_z &= \partial_y \partial_x \widetilde G_z
	\end{align}
\end{subequations}
which holds for all $\widetilde G_z$.

\section{Summary and conclusion}\label{sect-s}

A simple description of the dynamics of a near-resonant particle (time of flight $\approx$ optical period) traversing a DLA unit cell was presented. The description introduces
complex-valued functions
$\widetilde G_x(x,y), \widetilde G_y(x,y), \widetilde G_z(x,y)$ which are a natural generalization
of acceleration and deflection gradients used in previous DLA literature \cite{Plettner2007, PlettnerByer2008, LeedlePease2015, WeiJamison2017}.
The Panofsky-Wenzel theorem implies that $\widetilde G_x$ and $\widetilde G_y$
can be calculated from $\widetilde G_z$, so both the longitudinal and the transverse
dynamics can be derived from a single function $\widetilde G_z(x,y)$, the acceleration gradient.
This approach emphasizes a link between the previous DLA literature and the Panofsky-Wenzel theorem.

\section*{Acknowledgments}
I gratefully acknowledge receiving helpful remarks from Uwe Niedermayer, Steven Jamison, Joel England and Joshua McNeur.




%

\end{document}